\documentclass{article}

\usepackage{arxiv}%\usepackage{cite}
\usepackage{amsmath,amssymb,amsfonts}
\usepackage{algorithmicx,algpseudocode}
\usepackage{algorithm}
\usepackage{graphicx}
\usepackage{textcomp}
\usepackage{multirow}
\def\BibTeX{{\rm B\kern-.05em{\sc i\kern-.025em b}\kern-.08em
    T\kern-.1667em\lower.7ex\hbox{E}\kern-.125emX}}
\usepackage{float}

\usepackage{amsmath}
\usepackage{amssymb}
\usepackage{graphicx}
\usepackage{program}
\usepackage{ifthen}
\usepackage{epsfig}
\usepackage{array}
\usepackage{color}
\usepackage[table]{xcolor}
\usepackage{xspace}
\usepackage{url}
\usepackage{listings}
\usepackage{amssymb}% http://ctan.org/pkg/amssymb
\usepackage{pifont}% http://ctan.org/pkg/pifont

\usepackage{verbatim}
\usepackage{booktabs}
\usepackage[group-minimum-digits=4,group-separator={,}]{siunitx}

\usepackage{subcaption}

\definecolor{darkgreen}{rgb}{0,0.44,0}
\definecolor{darkred}{rgb}{0.44,0,0}
\definecolor{darkblue}{rgb}{0,0,0.44}

\definecolor{myred}{rgb}{0.82,0.0,0.0} % red
\definecolor{mygreen}{rgb}{0.0,0.44,0.0} % green
\definecolor{myblue}{rgb}{0.0,0.0,0.82} % blue
\definecolor{mygray}{rgb}{0.9,0.9,0.9}
\definecolor{mymauve}{rgb}{0.58,0,0.82}
\definecolor{mycreme}{rgb}{1.0,0.8,0.2} % creme

\newcommand{\crr}[1]{\textcolor{myred}{#1}}
\newcommand{\crb}[1]{\textcolor{myblue}{#1}}
\newcommand{\crg}[1]{\textcolor{mygreen}{#1}}

\newcommand{\pydl}{\textsf{PyDTNN}\xspace}
\newcommand{\imcol}{\textsc{im2col}\xspace}
\newcommand{\base}{\textsc{base}\xspace}
\newcommand{\conv}{\textsc{conv}\xspace}

\newcommand{\fuse}{\textsc{fuse}\xspace}

\newcommand{\cache}{\textsc{cache}\xspace}

\newcommand{\MAXN}{\textsf{MAXN}\xspace}
\newcommand{\WALL}{\textsf{30W ALL}\xspace}

\newcommand{\convgemmn}{\mbox{\sc convgemm}}
\newcommand{\convgemm}{\convgemmn\xspace}
\newcommand{\gemm}{\textsc{gemm}\xspace}%
\newcommand{\cython}{\textsc{cython}\xspace}%

\graphicspath{{Figures/}}

\usepackage{amssymb}
\title{High Performance and Energy Efficient Inference\\ for Deep Learning on %Multicore 
ARM Processors}

\author{ Adrián Castell\'o \\
	Universitat Polit\`ecnica de Val\`encia\\
	\texttt{adcastel@disca.upv.es} \\
	\And
	Sergio Barrachina \\
        Universitat Jaume I \\
	\texttt{barrachi@uji.es} \\
	\And
	Manuel F. Dolz \\
        Universitat Jaume I \\
	\texttt{dolzm@uji.es} \\
	\And
	Enrique S. Quintana-Ort\'i \\
	Universitat Polit\`ecnica de Val\`encia\\
	\texttt{quintana@disca.upv.es} \\
	\And
	Pau San Juan \\
	Universitat Polit\`ecnica de Val\`encia\\
	\texttt{p.sanjuan@upv.es} \\
}

\begin{document}

\maketitle

\begin{abstract}
We evolve \pydl, a framework for distributed parallel training of Deep Neural Networks (DNNs), into an efficient
inference tool for convolutional neural networks. Our optimization process  on multicore ARM processors
involves several high-level transformations of the original framework, such as
the development and integration of Cython routines to exploit thread-level parallelism;
the design and development of 
micro-kernels for the matrix multiplication, vectorized with ARM's NEON intrinsics, 
that can accommodate layer fusion;
and the appropriate selection of several cache configuration parameters tailored to the memory hierarchy of the
target ARM processors.

Our experiments evaluate both inference throughput (measured in processed images/s) and inference latency (i.e., time-to-response) as well as energy consumption per image when varying the level of thread parallelism and the processor power modes.
The experiments with the new inference engine are reported for the ResNet50 v1.5 model on the ImageNet dataset from the MLPerf suite using 
the ARM v8.2 cores in the NVIDIA Jetson AGX Xavier board. These results show 
superior performance compared with the well-spread TFLite 
from Google and slightly inferior results when compared with ArmNN, the native library from ARM for DNN inference.

\end{abstract}

%% Graphical abstract
%% \begin{graphicalabstract}
%% \includegraphics{grabs}
%% \end{graphicalabstract}

%% Research highlights
%% \begin{highlights}
%% \item \textcolor{red}{Pending}
%% \item Research highlight 2
%% \end{highlights}

%% \begin{keyword}
%% Deep learning, % \sep
%% inference, % \sep
%% matrix multiplication, % \sep
%% low-power multicore architectures, % \sep
%% high performance
%% \end{keyword}

%\end{frontmatter}

%% \linenumbers

%% main text
\section{Introduction}

Information technology companies are nowadays strongly interested in
running deep learning (DL) models at the edge to improve security (safety and privacy), to accelerate
the time-to-response (i.e., latency) experienced by the end-user,
and to reduce the energy consumption for IoT (Internet-of-Things) applications~\cite{8327042,park2018deep,8675201,Sha15}.
The deployment of these trained DL models, known as \textit{inference},
is often performed
% is preceded by
% a costly off-line training stage of one (or several candidates) complex
% deep neural networks  (DNNs),
% usually performed on a high-performance facility. % (a cluster or similar).
% This is then followed by a second stage, known as \textit{inference},
% that involves the deployment, tuning and usage of the  trained model,
on a wide variety of user appliances, ranging from drones and mobile phones to wearables and
IoT sensors~\cite{8327042,park2018deep,8675201}.
In this scenarios, the inference process is computationally less expensive than the
prior training stage, but often presents strict response time and/or energy
constraints and has to be performed on devices with limited computational and memory capacities,
as well as constraints in power supply.

% Depending on the specifications of the DL problem and target device,
% the inference stage may require the application of different techniques to the (DNN) model with the goal of
% reducing its memory occupancy and, to a lesser extent, %as well as %and %o a minor degree also
% computational complexity, power dissipation, and/or energy consumption.
% These techniques include pruning, quantization, weight sharing, compression of the model parameters,
% etc.~\cite{Han15}.
%Among them, quantization compresses the model by reducing the number of bits used to represent the model
%parameters (weights). As a result, for some DL problems, this clustering technique can result in accurate models that employ
%16 bits or less per weight and involve integer arithmetic only~\cite{Han15}.

In this paper, we investigate the efficient realization of an inference module for multicore ARM processors, making
the following contributions:
\begin{itemize}
\item Starting from the \pydl framework for distributed DNN training on clusters of computers~\cite{Barr21},
      we obtain a basic module for inference based on the forward pass stage of the original framework, enhanced with some preliminary optimizations
      that substitute a few key Python routines with efficient Cython-based counterparts parallelized with
      OpenMP directives.
\item For the convolution layers, we follow our work in~\cite{convgemm} to adopt a blocked variant of the \imcol
      transform~\cite{Che06} that casts this operation into a matrix multiplication (\gemm)
      while eliminating the high memory costs of its conventional formulation. This is achieved via
      careful utilization of the packing routines in the BLIS~\cite{BLIS1} realization of the matrix multiplication kernel.
\item We optimize the BLIS \gemm kernel with a dynamic mechanism to utilize architecture-specific cache
      configuration parameters. In addition, we expand the realization of \gemm in BLIS with a variant that
      targets the convolution operators to the most appropriate level of the cache.
\item We develop a new micro-kernel for the BLIS realization of \gemm using NEON vector intrinsics for ARM processors.
      This opens the door to fuse (i.e., merge) consecutive layers, reducing the overhead of memory access for some DNN models.
\item We conduct a complete, incremental evaluation of the impact of each one of these contributions, using a standard
      use case from the MLPerf benchmark suite (v1.0)\footnote{\url{https://mlperf.org/inference-overview}}
      for inference, on the ARM v8.2 Carmel processor embedded in the NVIDIA Jetson AGX Xavier board.
      Our experiments demonstrate that the resulting module offers competitive performance with state-of-the-art inference modules from ARM (ArmNN) and Google (TensorFlow Lite) while requiring only high-level transformations of the code.
\item Finally, we evaluate the performance of an alternative parallel scheme that prioritizes inference throughput versus latency and we include a complete study of the energy consumption.
\end{itemize}

The rest of the paper is structured as follows.
In Section~\ref{sec:opt-pydl} we review and evaluate the prototype inference module obtained by applying a few basic
optimizations to the code for the forward pass in \pydl.
In Section~\ref{sec:opt-conv} we describe several advanced optimizations introduced in this
basic module.
Next, in Section~\ref{sec:evaluation}, we
analyze other parallelization alternatives as well as energy consumption.
Finally, in Section~\ref{sec:remarks}, we close the paper with a few concluding remarks.

\section{Basic Module for DL Inference}
\label{sec:opt-pydl}

In contrast with the training stage, the inference process is considerably less
expensive though, when performed on edge devices, may present severe constraints in the response time, energy consumption,
and/or memory requirements. In this section we describe our initial work to adapt the \pydl framework to common inference scenarios, targeting low power ARM-based architectures.
%while maintaining most of the framework's appealing features regarding flexibility, interface, functionality, and performance.
To illustrate this process, we leverage the ResNet50 v1.5 model+ImageNet benchmark included in the MLPerf inference suite,
focusing on
the Carmel multicore processor (ARM v8.2) embedded in NVIDIA's Jetson AGX Xavier development kit.

\subsection{The \pydl framework for DL}

\pydl\footnote{The \pydl framework is available at \url{https://github.com/hpca-uji/PyDTNN/}, under a GNU General Public License v3.0.}
%(\textit{Python Distributed Training of Neural Networks})
is a framework for distributed training of DNNs on clusters of computers, written in Python, that
%was originally
is designed as a research-oriented tool with a low learning curve.
\pydl
%presents the following appealing properties:
%\begin{itemize}%{--}{}
%\itemsep 0.0cm
%\item \textbf{Flexible:} \pydl
prioritizes simplicity to facilitate that users can adapt the framework to prototype research exploration;
%\item \textbf{User-amiable interface:}
%\pydl
exposes an interface akin to  that of popular DL packages such as Keras; and
%\item \textbf{Ample functionality:} \pydl
supports a significant part of  common DNN models such as
multi-layer perceptrons (MLPs), convolutional neural networks (CNNs), residual networks
(ResNets), and transformers for natural language processing.
In addition,
%for this type of models,
\pydl
offers validation accuracy and parallel performance, for DL training, on par with those attained by Google's TensorFlow~\cite{Barr21}.
To attain this, \pydl leverages high performance computational and communication libraries such as, for example,
%when the target cluster is equipped with NVIDIA's GPUs, \pydl leverages
Intel's MKL; NVIDIA's \textsf{cuDNN}, \textsf{cuBLAS} and \textsf{NCCL}; 
as well as specialized implementations of MPI.
%\item \textbf{High performance:}
%\pydl leverages \textit{data parallelism} (DP)~\cite{DBLP:journals/corr/abs-1802-09941},
%via specialized message-passing libraries and kernels from high performance multi-threaded libraries
%for the major computational operations in CPU and GPUs.
%%In particular, when the target cluster is equipped with NVIDIA's GPUs, \pydl leverages
%%\textsf{cuDNN}, \textsf{cuBLAS}, and \textsf{NCCL}
%On cluster of computers equipped with graphics accelerators, \pydl delivers  parallel performance that is
%competitive with that of TensorFlow running on top of Horovod.
%\end{itemize}%{}
st

%\pydl is mostly written in Python, generating one process (MPI ranks) per cluster node to extract parallelism on distributed
%platforms.
%This is appropriate for DNN training, which is a computationally costly stage yet can
%be performed off-line.
%%In particular,

%As argued in the previous section, the \pydl framework was originally designed
%as an off-line prototyping tool, for large-scale DNN training,  on clusters of computers.
%With that consideration, the development
%\pydl was originally focused on distributed DNN training,
%prioritizing simplicity and flexibility over raw performance.

\subsection{ResNet50 v1.5 and ImageNet}

CNNs are especially appropriate DL technologies for image recognition, recommendation systems, image classification,
medical image analysis, natural language processing, brain-computer interfaces, financial time series, etc.
The importance of CNNs is recognized by the MLPerf suite v1.0,
which includes the ResNet50 v1.5 CNN model combined with the ImageNet (224$\times$224) dataset
as one of its six benchmarks.
%In the ResNet50 v1.5 model we can highlight
This model consists of 176 layers, comprising a large number of four types of \textit{transforms}:
53 convolutions, 48 non-linear (ReLU) functions, 52 batch normalizations, and 2 pooling operators;
see \cite{goyal2017accurate} for details.

\subsection{Jetson AGX Xavier}

%The experiments in this work %Sections~\ref{sec:opt-pydl} and~\ref{sec:opt-conv}
%were carried out on an
The NVIDIA Xavier board embeds an ARM Carmel 8-core CPU
%(implementing the ARMv8.2 micro-architecture),
(with the ARMv8.2-FP16 extension),
an NVIDIA 512-CUDA core Volta GPU, and \SI{32}{GiB} of main memory.
(As the target of this work is the optimization of inference on ARM architectures, we will not consider the GPU
hereafter.)
On the software side, the board runs under %experiments utilized
the Ubuntu Linux distribution 18.04.4, and
includes the GNU C compiler gcc 10.0 and BLIS 0.7.0 for the linear algebra kernels.

% The NVIDIA Carmel processor consists of 4 core clusters each with 2 ARM~v8.2 cores and 2MB 16-way set-associative of L2 data cache shared between them. Each core owns a \SI{64}{KiB} 4-way set-associative L1 data cache. All the core clusters are connected to the system fabric which provides coherence between the core clusters and the rest of the system. The coherence fabric also connects to a \SI{4}{MiB} 16-way set-associative victim-style L3 cache.
% The peak performance of this processor is XXX GFLOPS (billions of floating-point operations per second) when all
% 8 cores operate at XX~GHz.

For the experiments in this section, we employ all 8 ARM-based Carmel cores of the platform.
%and a batch size $t$=128.
Furthermore,
%As our main interest is testing the inference realization of \pydl
%in low-power systems,
we set the \MAXN power mode available
in the \emph{nvpmodel} power/performance management utility included in the NVIDIA Jetson AGX Xavier. This mode activates
all 8 Carmel cores and sets their frequency to \SI{2.3}{GHz}. % to ensure that the power consumption of the system stays within a \SI{30}{watts} budget.
This permits an easier evaluation of the multi-threaded parallelization,
as it avoids side effects that would occur if the system
was allowed to automatically adjust the core frequencies depending on the workload and the number of active cores.

All tests in the paper employ IEEE 32-bit floating-point arithmetic (FP32). Furthermore,
the tests were executed multiple times and the results were averaged to smooth out system load effects in the measurements.

\subsection{Baseline inference}

As a starting point for our work, we obtained a baseline module for inference (hereafter referred to as \base) by simply utilizing the  Python routine in \pydl for the training \textit{forward pass}~\cite{8114708,Pouyanfar:2018:SDL:3271482.3234150}.
This prototype module presents the following features:
\begin{list}{--}{}
 \itemsep=0pt
 \item The convolutions are cast in terms of \gemm kernels via an \imcol re-organization of the activation inputs that constructs a large augmented matrix before invoking \gemm~\cite{Che06}.
 \item The \gemm operation is realized via \texttt{NumPy}, which is linked against the ARM-optimized realization of this kernel in the BLIS framework~\cite{BLIS1}. This linear algebra library leverages multi-threaded parallelism using OpenMP and exploits the vector units in the ARM processor via an assembly-encoded micro-kernel with vector instructions. (BLIS is described in more detail in Section~\ref{sec:opt-conv}.)
 \item The batch normalizations are implemented as Python routines.
 \item The elementwise ReLU functions are encoded as Cython routines parallelized with OpenMP.
 \item The pooling layers are implemented as Cython routines that parallelize the \imcol transform using OpenMP directives.
\end{list}

We first offer an analysis of the time costs of the inference process using the prototype inference module;
see the column labeled as \base in Table~\ref{tab:cost}.
Given the large number of transforms comprised by the ResNet50 v1.5 model (more than 150),
we group the costs in the table into the four main
types of transforms: 2D convolutions, batch normalizations, ReLU activation functions, and pooling operators.
The costs reported there correspond to the seconds that are necessary to process a batch consisting of $t$=128 images and the global throughput is also measured in number of images per second. % ($=t$/time).

\newcommand{\cbase}{\multicolumn{2}{c|}{Basic module (Section~\ref{sec:opt-pydl})}}
\newcommand{\coptm}{\multicolumn{3}{c}{\gemm optimizations (Section~\ref{sec:opt-conv})}}
\begin{table}[tbh]
 \centering
 %\resizebox{\linewidth}{!}{
 \begin{tabular}{l|rr|rrr}
  \toprule
  Type of   & \cbase    & \coptm                                         \\
  transform & \base{}   & \cython{} & \conv-opt & \cache-opt & \fuse{}   \\ \midrule
  Conv2D    & 10.81     & 10.81     & ~8.37     & ~7.92      & ~7.78     \\
            & (44.82\%) & (81.34\%) & (77.14\%) & (78.49\%)  & (87.21\%) \\ [0.2in]
  Batch     & 10.58     & ~0.55     & ~0.55     & ~0.55      & --        \\
  norm.     & (43.86\%) & ~(4.14\%) & ~(5.07\%) & ~(5.45\%)  &           \\ [0.2in]
  ReLU      & ~1.36     & ~1.36     & ~1.36     & ~1.36      & ~0.75     \\
            & ~(5.64\%) & (10.23\%) & (12.53\%) & (13.47\%)  & ~(8.41\%) \\ [0.2in]
  Pooling   & ~1.08     & ~0.26     & ~0.26     & ~0.26      & ~0.26     \\
            & ~(4.48\%) & ~(1.96\%) & ~(2.40\%) & ~(2.57\%)  & ~(2.91\%) \\ \midrule
  \footnotesize{\textbf{Time (s)}}
            & 24.08     & 14.42     & 10.85     & 10.09      & ~8.92     \\
  \footnotesize{\textbf{Images/s}}
            & ~5.31     & ~8.87     & 11.80     & 12.68      & 14.34     \\ \bottomrule
 \end{tabular}
 %}
 \caption{Cost analysis of the inference variants derived from \pydl{} when applied to ResNet50 v1.5+ImageNet and batch size $t$=128 using the full 8 Carmel cores of NVIDIA's Jetson AGX Xavier.}
 \label{tab:cost}
\end{table}

CNNs avoid overfitting by taking advantage of the hierarchical structure of the data.
This is achieved via convolutional layers,
which in general concentrate a significant fraction of the computational cost for CNNs.
This expectation is confirmed only partially by the results in Table~\ref{tab:cost},
which show that the convolutions
consume 44.82\% of the execution time of the initial
\base module. However, the Python realization of the batch normalizations
is almost as expensive (43.86\%).
From this analysis, it is clear that, for this particular testbed (ResNet50 v1.5 with ImageNet and NVIDIA's Carmel processor),
we should consider the convolutions and batch normalizations as the two first targets in the optimization of the baseline inference module.
% For the particular case of the convolution, based on an \imcol re-organization of the activation inputs followed by
% a \gemm, we will first assess the overhead introduced by the former.
% For this purpose, we consider
% the number of floating-point operations (flops)
% necessary for each transform divided by the time cost to perform these as well as the data movements incurred by
% the application of \imcol.
% \textcolor{red}{Illustrate behaviour a plot with performance for \gemm only, \imcol+\gemm and \convgemm.}

%% \textcolor{red}{%
%% After the analysis of the baseline, organize the following sections in this order:
%% \begin{enumerate}
%% \item First: batch norm and maxpool optimizations (together).
%% \item Second: Convolutions via \convgemm or \imcol.
%% \item Performance optimization of BLIS via cache configuration parameters: $m_c,n_c,k_c$ and 8x8 micro-kernel with
%% prefetching+NEON intrinsics.
%% \item Layer fusion inside micro-kernel.
%\end{enumerate}
%}

\subsection{Batch normalization}

A primary optimization target corresponds to the batch normalizations,
which, surprisingly, were responsible for almost 44\% of the execution time in the \base variant.
%This is quite surprising, given the theoretically low-arithmetic cost of this type of calculation.
An inspection of the Python code for the realization of this transform in \pydl, together with some additional experiments,
guided us to conduct the following optimizations:
\begin{list}{--}{}
 \itemsep=0pt
 \item Elimination of code invariants in this type of transforms, in particular, the calculation of the standard deviation and mean, which can be directly obtained from the prior training process. (This unnecessary recalculation was due to this routine being used for the training forward pass in \pydl, where these two parameters must be computed for each new batch.)
 \item Replacement of the Python code for this calculation with a Cython routine that is parallelized (via OpenMP) and vectorized (using the appropriate compiler directives).
 \item Avoidance of unnecessary accesses to large data arrays thanks to a more careful use of temporary variables.
 \item Adoption of column-major storage for the data arrays that conform to the accesses to these structures to favor a more efficient, vectorized data retrieval.
\end{list}

\begin{figure}[tbh!]
 \centering\vspace{0.1cm}
 \includegraphics[width=\textwidth]{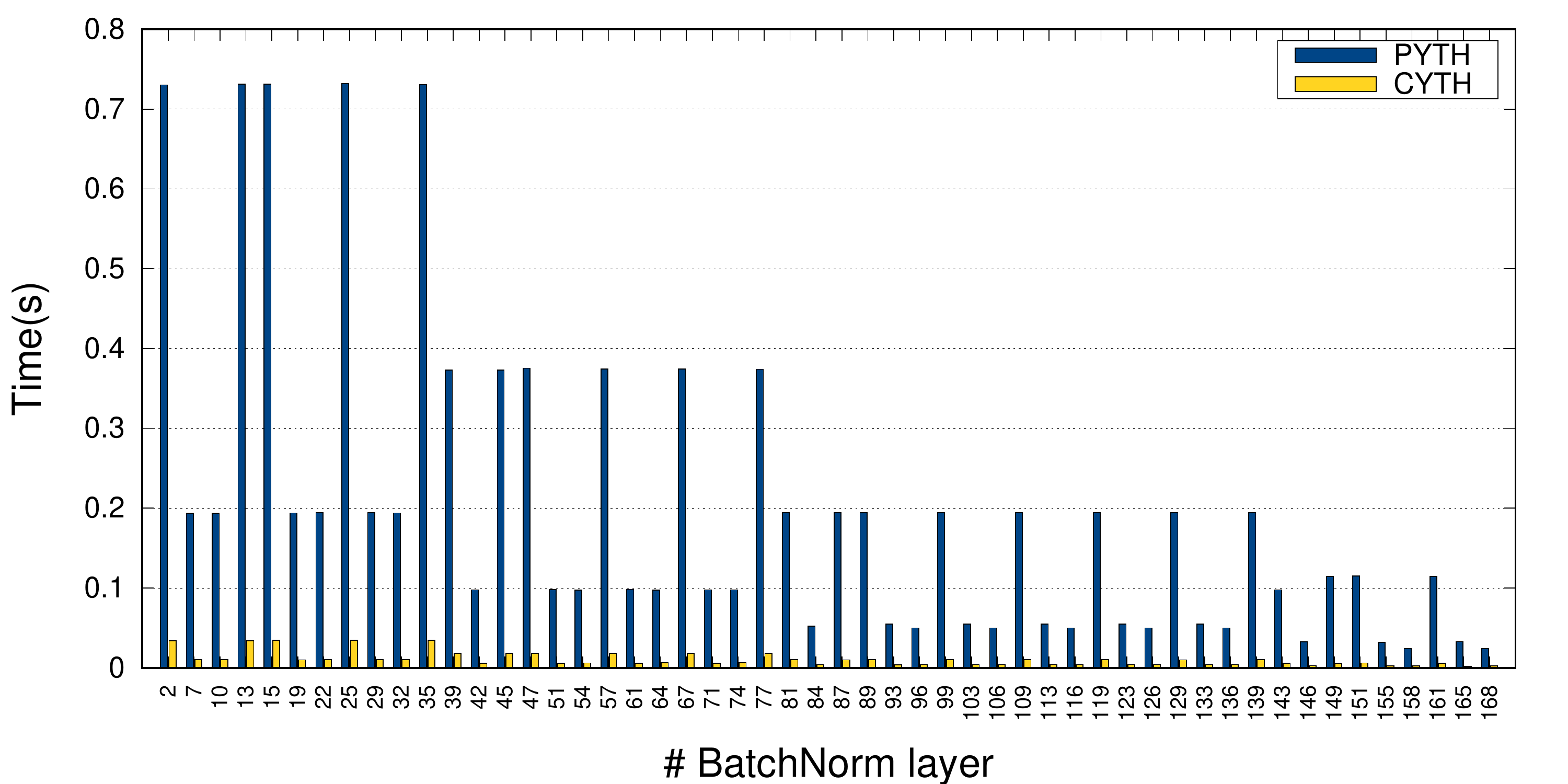}
 \caption{Performance of the two options for batch normalization when applied to
  ResNet50 v1.5+ImageNet and batch size $t$=128 using the full 8 Carmel cores of NVIDIA's Jetson AGX Xavier.}
 \label{fig:performance_batchnorm}
\end{figure}

Figure~\ref{fig:performance_batchnorm}
illustrates the considerable benefits attained by the optimized realization
versus the original one
(respectively labeled as \textsf{PYTH} and \textsf{CYTH} in the figure).
These results motivated us to apply a similar re-formulation of the Cython-based pooling transforms
in the ResNet50 v1.5 model, which were responsible
for 4.48\% of the total cost in the \base module.
(There exists a second pooling transform in the final layers model, but its cost is negligible.)
%The optimization, in this case, consists in the development of a multi-threaded and
%vectorized Cython version of the original Python code that, in addition, manually unrolls some of the
%loops to avoid the overhead of iterating over small loops.
%As a result, the cost of the pooling is reduced from \SI{0.80}{s} to \SI{0.20}{s}, and the throughput rate is raised
%from \SI{13.25}{images/s} to the final \SI{14.12}{images/s}.

The positive impact of these optimizations is confirmed in the column labeled as \cython in Table~\ref{tab:cost}, which shows that the time spent for the batch normalizations and pooling transforms is reduced from \SI{10.58}{s} and \SI{1.08}{s} in the \base module to \SI{0.55}{s} and \SI{0.26}{s}, respectively, in the \cython{} module.
As a result, in the new \cython variant, the batch normalization and pooling transforms contribute with much lower costs to the total execution time: 4.14\% and 1.96\%, respectively. %for variant \cython.
The global outcome of this optimization is an acceleration of the processing throughput from \SI{5.31}{images/s} for the former to \SI{9.63}{images/s} for the latter (a speedup of 1.81).
%and the  ReLU function emerges as a relevant optimization objective.

\section{Optimization of GEMM-based Convolutions for Inference}
\label{sec:opt-conv}

After the initial optimization proposed in the \cython variant,
the previous section identified the convolution as the key contributor to the practical cost of the
inference module when applied to ResNet50 v1.5 in the target Jetson AGX Xavier platform.
In this section, we describe some ARM architecture-aware
optimization techniques that significantly reduce the cost of this operator. Prior to this, we open the section with a
short review of the BLIS approach to obtain a portable, high-performance realization of \gemm.

\subsection{BLIS: Open and portable kernels for dense linear algebra}
\label{subsec:blisgemm}

Consider the \gemm operation
$C \mathrel{+}= A \cdot B$, where
$C \rightarrow m \times n$, $A \rightarrow m \times k$, and $B \rightarrow k \times n$.
BLIS mimics GotoBLAS~\cite{Goto:2008:AHP}
to implement this kernel  as three nested loops around a \textit{macro-kernel} plus two \textit{packing routines}.
From that point, BLIS differs from GotoBLAS by decomposing the macro-kernel to expose two more loops around a micro-kernel;
see~\figurename~\ref{fig:gotoblas_gemm}
and~\cite{BLIS1} for details.

\begin{figure*}[t]
 \centering
 \begin{minipage}[c]{\textwidth}
  \resizebox{\linewidth}{!}{
   \begin{tabular}{llll}
    \textsf{L1:} & {\bf for} $j_c$ = $0,\ldots,n-1$ {\bf in steps of} $n_c$                                                                                \\
    \textsf{L2:} & \hspace{4ex}  {\bf for} $p_c$ = $0,\ldots,k-1$ {\bf in steps of} $k_c$                                                                  \\
                 & \hspace{8ex}           \crb{$B(p_c:p_c+k_c-1,j_c:j_c+n_c-1)$} $\rightarrow \crb{B_c}$            &                 & // Pack into $B_c$ \\
    \textsf{L3:} & \hspace{8ex}           {\bf for} $i_c$ = $0,\ldots,m-1$ {\bf in steps of} $m_c$                                                         \\
                 & \hspace{12ex}                     \crr{$A(i_c:i_c+m_c-1,p_c:p_c+k_c-1)$} $\rightarrow \crr{A_c}$ &                 & // Pack into $A_c$ \\
    %&\hspace{12ex} \crg{$C(i_c:i_c+m_c-1,j_c:j_c+n_c-1)$}
    %~$\mathrel{+}=$ ~\crr{$A_c$}
    %~$\cdot$\!~\crb{$B_c$}  & & // Macro-kernel\\
    \cline{2-4}
    \textsf{L4:} & \hspace{12ex} {\bf for} $j_r$ = $0,\ldots,n_c-1$ {\bf in steps of} $n_r$                         &                 & // Macro-kernel    \\
    \textsf{L5:} & \hspace{16ex}   {\bf for} $i_r$ = $0,\ldots,m_c-1$ {\bf in steps of} $m_r$                                                              \\
    \cline{2-3}
                 & \hspace{20ex}             \crg{$C_c(i_r:i_r+m_r-1,j_r:j_r+n_r-1)$}                               & // Micro-kernel                      \\
                 & \hspace{24ex} ~$\mathrel{+}=$     ~\crr{$A_c(i_r:i_r+m_r-1,0:k_c-1)$}                                                                   \\
                 & \hspace{24ex} ~~~$\cdot$\!~~~~\crb{$B_c(0:k_c-1,j_r:j_r+n_r-1)$}                                                                        \\
   \end{tabular}
  }
 \end{minipage}
 %\end{tabular}
 \caption{High performance implementation of \gemm in BLIS.
  $C_c \equiv C(i_c:i_c+m_c-1,j_c:j_c+n_c-1)$
  is a notation artifact, introduced to ease the presentation of the algorithm.
  In contrast, $A_c,B_c$ denote buffers that are involved in data copies.
  For simplicity, we consider that $m,n,k$ are integer multiples of $m_c,n_c,k_c$ respectively,
  and $m_c,n_c$ are integer multiples of $m_r,n_r$ respectively.}
 \label{fig:gotoblas_gemm}
\end{figure*}

\begin{figure}[t]
 \centering
 \begin{tabular}{c}
  \includegraphics[width=0.7\textwidth]{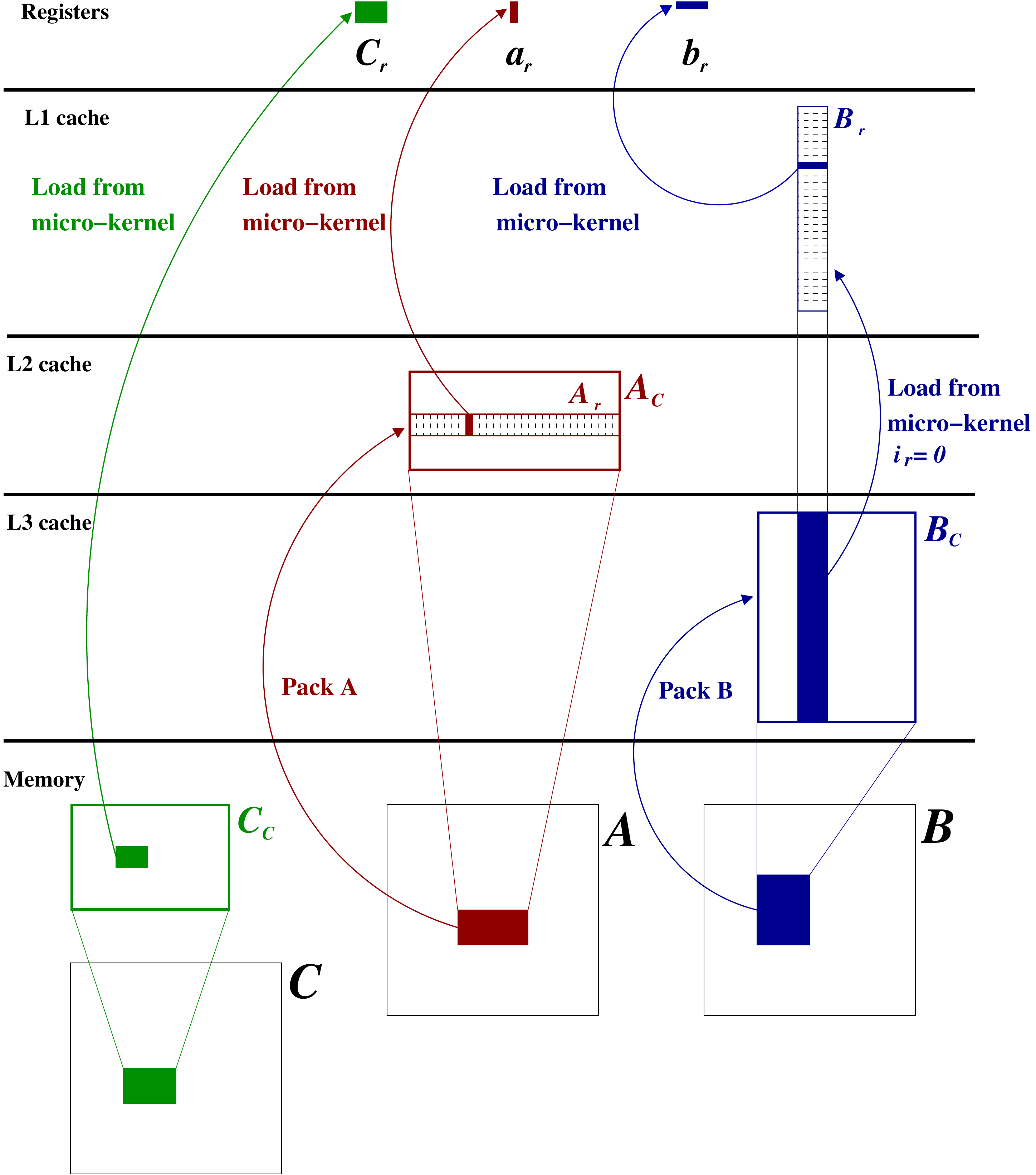}
 \end{tabular}
 \caption{Data movement in the BLIS implementation of \gemm.}
 \label{fig:blis_movement}
\end{figure}

The architecture-specific optimization of the BLIS kernel requires a selection of the loop strides
$m_c,n_c,k_c$ that matches the cache organization of the target processor~\cite{BLIS4}
(plus the development of a vectorized version of the micro-kernel, to be discussed later).
In some detail,  the loop ordering in the BLIS realization of \gemm, together with the packing
routines, and a proper selection of the cache configuration parameters/loop strides ($n_c$, $k_c$, $m_c$),
%that matches the processor memory hierarchy,
orchestrate a regular pattern of data transfers,
as illustrated in Figure~\ref{fig:blis_movement}.
Concretely, packing $B$ into $B_c$ inside loop~\textsf{L2} of the BLIS kernel
in Figure~\ref{fig:gotoblas_gemm} makes a copy of these data into the L3 cache,
and the re-use of this particular buffer for all iterations of the subsequent loop,
\textsf{L3}, favors that this buffer persists in that level of the cache. Similarly, packing
$A$ into $A_c$ in Loop~\textsf{L3} copies these data into the L2 cache and the repeated access
to this buffer in loop~\textsf{L4} preserves in that level.
For a detailed discussion, see~\cite{BLIS1,BLIS4}.

The development of an NVIDIA Carmel-specific micro-kernel
using assembly code was done as part of the work in~\cite{Catalan2016}.
For a multi-threaded execution of the BLIS realization,
the loops of the \gemm kernel to be parallelized can be selected at execution time.
The OpenMP-based parallelization of the BLIS \gemm kernel has been previously analyzed for conventional multicore processors~\cite{BLIS2}, modern many-threaded architectures~\cite{BLIS3},
and low-power (asymmetric) ARM-based processors in~\cite{Catalan2016}.

\subsection{Convolution via \convgemm}

One major goal of packing in BLIS is  to arrange the entries of
$A$ and $B$ into $A_c$ and $B_c$, respectively,
so that the elements of these buffers
are accessed with unit stride when executing the micro-kernel~\cite{BLIS1}.
In practice, provided $m$ is large enough,
the cost of the packing for $B_c$ is negligible compared with the number of flops
performed inside Loop~\textsf{L3}.
(A similar reasoning applies to the overhead due to the packing for $A_c$ and the value of $n_c$.)

In~\cite{convgemm} we integrated the convolution within the BLIS~\cite{BLIS1} realization of \gemm,
obtaining a \convgemm routine that reduces the considerable memory requirements of the full \imcol transform.
To attain this, the \convgemm realization assembles the augmented activation matrix \textit{by blocks}
with the novelty that, for performance reasons, the block dimensions are adjusted
to the internal buffers utilized by BLIS \gemm to avoid the usage of extra memory while
efficiently accommodating the data in the processor cache hierarchy; see~\cite{convgemm} for the full details.

\begin{figure}[tbh]
 \centering\vspace{0.1cm}
 \includegraphics[width=\textwidth]{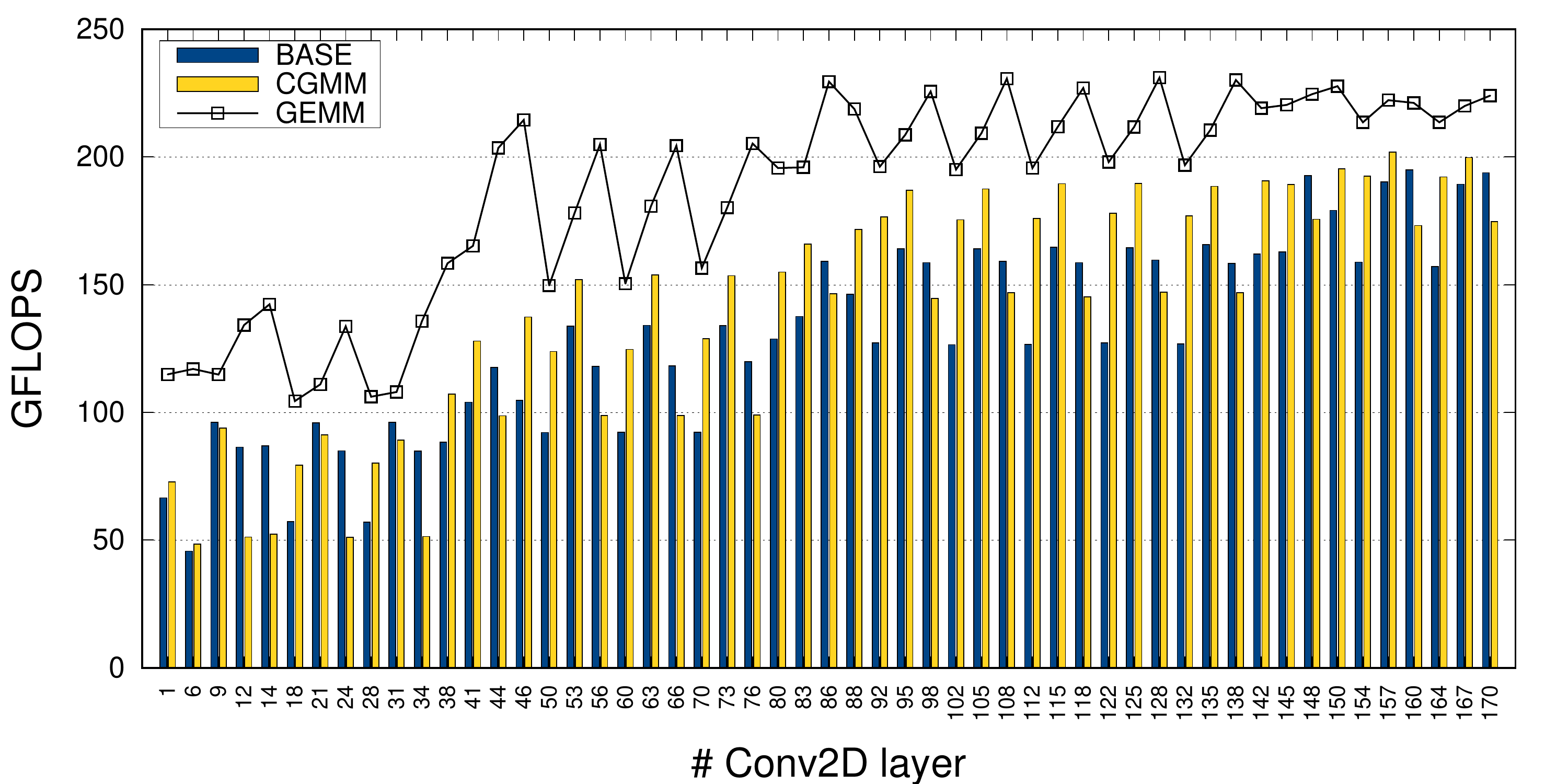}
 \caption{Performance of the two options for the convolution when applied to
  ResNet50 v1.5+ImageNet and batch size $t$=128 using the full 8 Carmel cores of NVIDIA's Jetson AGX Xavier.}
 \label{fig:performance_conv2d}
\end{figure}

Figure~\ref{fig:performance_conv2d} displays the impact on performance when using the \convgemm routine versus the full-\imcol approach for the convolutions appearing in the testbed that is targeted in this section. Performance is measured there in terms of billions of floating-point operations, abbreviated as flops, per second (GFLOPS), taking intone account the dimensions of the \gemm that is computed at each layer. For the two realizations of the convolution evaluated in the figure, \textsf{IM2COL+GEMM} employs the full \imcol transform followed by an invocation to \gemm whereas \textsf{CONVGEMM} directly calls the \convgemm routine that integrates the memory-saving blockwise variant of \imcol.
%\textcolor{red}{Discuss reduction in memory.}
In addition, for each layer, we include the GFLOPS attained by a direct invocation to \gemm
(with operands of the same dimensions) that
does not perform any type of \imcol transform. This last rate offers an estimation of the overhead present in the realizations based on the full- and block-wise \imcol.
In an independent experiment, we could confirm
that, in the latter case, this overhead is
mostly due to some data re-organizations which are necessary to call \convgemm, not to the block-wise \imcol itself.

The results in
Figure~\ref{fig:performance_conv2d} expose
that the best realization largely depends on the layer specifications (number of channels, number of filters, filter size,
strides, etc.).
In general, those
transforms for which both the kernel size and the number of channels are small
(e.g., 1$\times$1 kernels and 128 output channels), tend to favour
\textsf{IM2COL+GEMM} while in the remaining cases
\textsf{CONVGEMM} should be preferred. To accommodate this, in variant \conv-opt we fix the realization to utilize the most
efficient option on a per-transform (layer) basis.

The column labeled as \conv-opt in Table~\ref{tab:cost} reports the performance of this second variant, which
employs the best realization (\textsf{IM2COL+GEMM} or direct \textsf{CONVGEMM}) for each convolution transform
of ResNet50 v1.5, showing a considerable increase in the processing throughput with respect to the
\cython variant, from 9.63 to \SI{11.80}{images/s}. This represents a speed-up of 1.22.
%% The most significant difference between the two realizations is that \convgemm avoids
%% the assembly of the full augmented matrix,  which allows operating with larger batch sizes.
%% \textcolor{red}{Give explicit memory-saving numbers!}

\subsection{Optimization of cache usage for \gemm}

For performance and portability reasons, the values for the cache configuration parameters $n_c,k_c,m_c$ should be adjusted to the target processor cache hierarchy. This is done, in general,  as part of an offline optimization process that prioritizes performance for moderate to large, ``squarish'' problems, with $m=n=k \in O(10^3)$. For the NVIDIA Carmel processor and FP32 arithmetic, this experimental optimization process in BLIS yields a selection of $m_c=560, n_c=3,072, k_c=368$. These values ensure that $A_c$ fits into the L2 cache, and a panel of $B_c$ fits into the L1 cache. Furthermore, taking into account the associativity of these two cache levels, this selection aims to reduce the risk of evicting them from the corresponding level during the access to other data~\cite{BLIS4}. (At this point, we recall that the NVIDIA Carmel processor does not have an L3 cache.)

Unfortunately, many of the \gemm operations that are associated with the convolutions appearing in practical CNNs are
far from presenting such ``ideal'' dimensions. Concretely, when tackled via the \imcol transform, most of the initial convolutions
in the target ResNet50 v1.5 model involve
a matrix multiplication where $m,k$ are small (and equal), usually in $\{64,128,256\}$, while $n$ is much
larger, of $O(10^4-10^5)$.
The result is a suboptimal utilization of the cache memories and, in consequence, low performance. For example,
for the matrix multiplication in the first convolution of ResNet50 v1.5,
$m=k=64$ while $n \approx 140K$. Therefore, $A_c$ is a small $64 \times 64$ matrix, which occupies only a minor fraction of
the target 2-MiB L2 cache of the NVIDIA Carmel processor: Indeed, less than 1\%!

%\textcolor{red}{Pending: Discuss the selection of the convolution-aware cache configuration parameters}
To tackle this problem, we implemented our realization of the \gemm kernel, which follows the BLIS cache optimization
principles, but allows a dynamic selection of
$n_c,m_c,k_c$, at execution time, depending on the dimensions of the matrix multiplication appearing in each particular layer. In addition,
we implemented a variant of the BLIS kernel that ``swaps'' the target cache levels for $A_c$ and $B_c$
(see Figure~\ref{fig:blis_movement}), by interchanging loop
\textsf{L1} with
\textsf{L3}, and
loop \textsf{L4} with
\textsf{L5} of the BLIS kernel. This favors that a panel of
$A_c$ resides in the L1 cache while $B_c$ lies in the L2 cache. The purpose of this variant is to maximize the
benefits of accessing data in the L2 cache.
Finally,
we performed
an extensive experimental analysis to select the optimal values of these cache configuration parameters for the NVIDIA Carmel processor;
and fixed the \pydl-based inference module to utilize the most efficient option on a per-transform (layer) basis.

\begin{figure}[tbh!]
 \centering\vspace{0.1cm}
 \includegraphics[width=\textwidth]{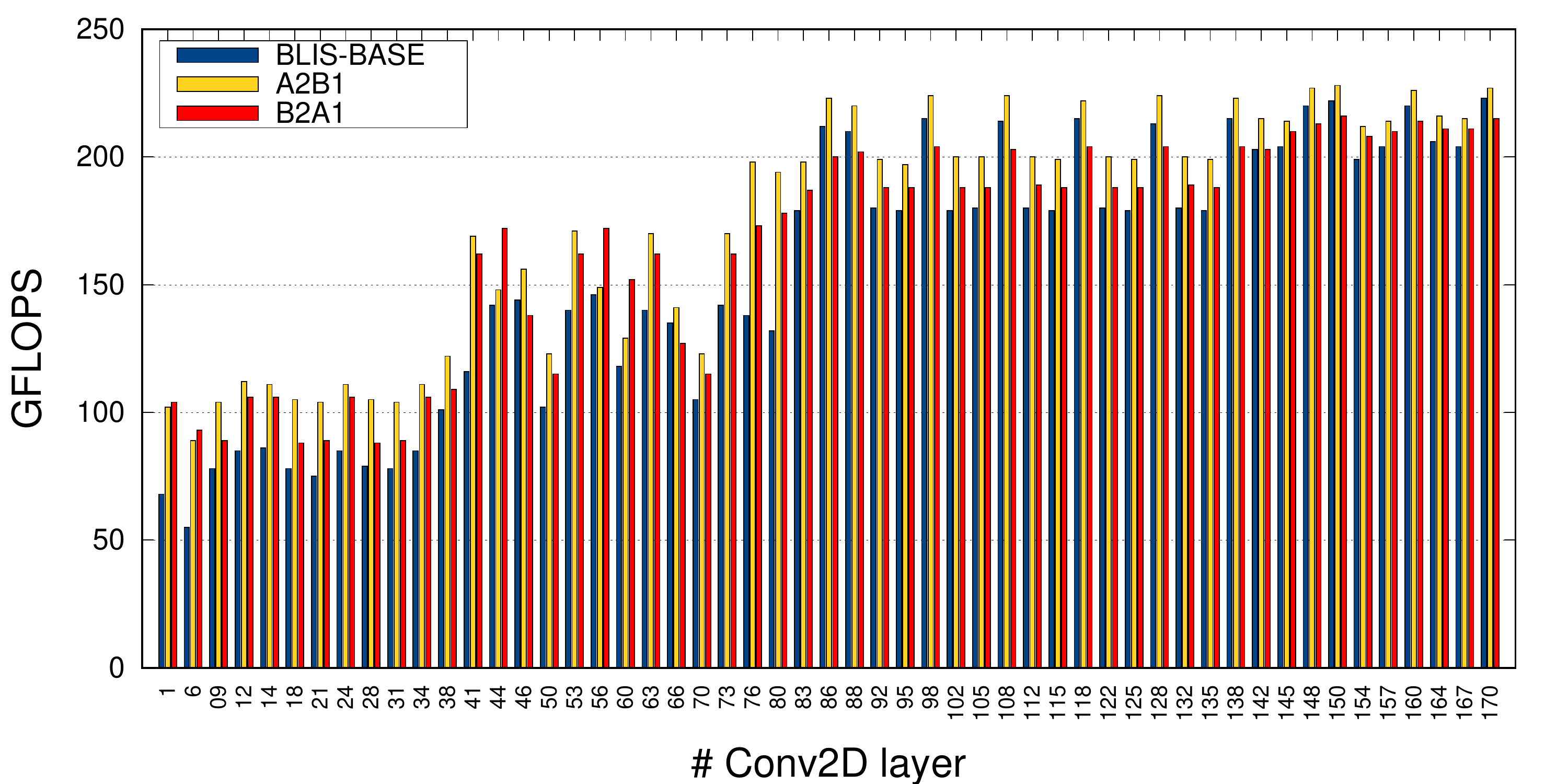}
 \caption{Performance of the matrix multiplications appearing in the convolution layers of
  ResNet50 v1.5+ImageNet with batch size $t$=128 using the full 8 Carmel cores of NVIDIA's Jetson AGX Xavier.}
 \label{fig:performance_conv2d_optgemm}
\end{figure}

Figure~\ref{fig:performance_conv2d_optgemm} reports the performance, in GFLOPS, of the original version of BLIS, labeled as
\textsf{BASE} there, compared with those of the two new variants that dynamically adjust the cache parameters taking into account
the parameters of each layer, with
$A_c$ in the L2 cache
and
(part of) $B_c$ in the L1 cache
(as in the original version of BLIS) or vice-versa. We identify these two
variants in the figure as
\textsf{A2B1} and
\textsf{B2A1}, respectively.
In general, we observe that the best variant is highly dependent on the layer characteristics, with
\textsf{A2B1} offering a better option in more cases. The results also demonstrate that a careful selection of the
variant outperforms the BLIS default option
(\textsf{BASE}) by a visible margin.
The largest gain is observed for the second layer, where variant
\textsf{B2A1} outperforms
\textsf{BASE} by 69\%, while the smallest gain appears in the penultimate layer, with a gain of less than 2\%. On average, the
gain is 18\% (arithmetic mean) and the weighted average gain is 21\%.

The global effect of the cache optimization techniques on the inference module is reported
in the column labeled as \cache-opt of Table~\ref{tab:cost}.
The results for the grouped Conv2D layers show a reduction of execution time from \SI{8.37}{s} for
\conv-opt to \SI{7.92}{s} for \cache-opt. This is lower than we could have estimated from the gains for \gemm due to the cache optimizations.
To explain this, we note that the acceleration reported
in Figure~\ref{fig:performance_conv2d_optgemm} was observed
for a standalone execution of the matrix multiplication kernel, that
is, without the data reorganization that is necessary for the preparation of the \imcol transform.
In contrast,
when these complete data transforms are taken into account, the overall gain due to this optimization is smaller,
yielding a speedup of 1.07, and an increase in the throughput rate from 11.80 to \SI{12.68}{images/s}.
%\textcolor{red}{Evaluate and comment impact of this optimization on inference! Refer to Table~\ref{tab:cost}.}

\subsection{High-level micro-kernels for the NVIDIA Carmel processor}

Let us now turn our attention to the micro-kernel.
As shown in Figure~\ref{fig:gotoblas_gemm},
this operation is responsible for performing the tiny matrix multiplication
$C_r \mathrel{+}= A_r \cdot B_r$, where
$C_r =  C_c(i_r:i_r+m_r-1,j_r:j_r+n_r-1) \rightarrow m_r \times n_r$;
$A_r =  A_c(i_r:i_r+m_r-1,0:k_c-1) \rightarrow m_r \times k_c$; and
$B_r =  B_c(0:k_c-1,j_r:j_r+n_r-1) \rightarrow k_c \times n_r$.
In practice,
the micro-kernel is implemented as a simple loop along the $k_c$  dimension of the product that updates $C_r$ with the outer
product of one column of $A_r$ and one row of $B_r$ per iteration:

\begin{center}
 \begin{minipage}[c]{0.7\textwidth}
  %\footnotesize
  \resizebox{\linewidth}{!}{
   \begin{tabular}{llll}
     & \hspace{0ex} {\bf for} $k$ = $0,\ldots,k_c-1$ {\bf in steps of} $1$                       & ~~~~ & Micro-kernel \\
     & \hspace{4ex}         \crg{$C_r$} $\mathrel{+}=$ \crr{$A_r(:,k)$}~$\cdot$ \crb{$B_r(k,:)$}
   \end{tabular}
  }
 \end{minipage}
\end{center}

All routines of the BLIS-inspired realization of \gemm are encoded in plain C except for the micro-kernel.
This enhances portability as migrating the kernel to a particular processor architecture
only needs to develop an efficient realization of that component for the target processor.
The micro-kernel
is usually vectorized using either architecture-dependent assembly instructions or vector intrinsics~\cite{BLIS1}.
As a rule of thumb,
$m_r,n_r$ are selected so that $C_r$, a column of $A_r$, and a row of $B_r$ occupy a significant fraction of the
processor (vector) registers. Besides, $k_c$ is set so that the cost of loading $C_r$ into the processor registers
(before the loop commences) and writing back the updated block into the main memory (once the loop is completed)
is amortized over a sufficient amount of flops; see~\cite{BLIS4}.
These conditions usually imply that $m_r,n_r \in O(10)$ and $k_c \in O(100)$.

For the ARM-based processors targeted in our work,
we developed several implementations of the micro-kernel using ARM NEON intrinsics. The best results were in general obtained
for a micro-kernel with $m_r \times n_r = 8 \times 8$, which loads the full $C_r$ into
16 vector registers (with 4 FP32 elements per 128-bit vector register).
In addition, the implementation integrates data pre-fetching~\cite{Hennessee} by
loading the first column/row of $A_r/B_r$ into 2+2 vector registers before the micro-kernel loop commences.
Then, at each iteration of the loop, say $k$,
the micro-kernel prefetches the $(k+1)$-th column/row of $A_r/B_r$ into 2+2 additional vector registers,
to then proceed to accumulate in $C_r$ the product involving the $k$-th column/row using vector instructions.

Our NEON-based implementation of the micro-kernel compiled with GNU \texttt{gcc-10} and the appropriate optimization flags
delivered a sustained performance that is similar to that of the original assembly-encoded micro-kernel in BLIS for ARM architectures.

The key observation though is that a ``high-level'' implementation of the micro-kernel, using plain C code plus ARM NEON intrinsics, paves the road to an efficient fusion of different types of layers, as described in the following subsection.

\subsection{Layer fusion}

The ReLU function is a simple test-and-set operator that is applied element-wise to the activations of a layer,
setting to zero all negative values and leaving the remaining ones unchanged.
The batch normalization involves a couple of shift and scale arithmetic operations involving mean and standard deviation parameters.
From the computational point of view, the arithmetic cost of these two types of transforms is low
compared with the convolution itself.
Therefore, their contribution to the total time in Table~\ref{tab:cost} seems excessive, especially when considering
the Cythonized, parallelized, and vectorized version of the kernels.

Some additional experiments allowed us to identify that these costs are mostly due to memory accesses. To tackle this, when possible,
we fuse (i.e., merge) the application of the batch normalization and ReLU with a previous convolution
into a single ``multi-layer operation.''
For this purpose, we take advantage of the high-level implementation of the micro-kernel
(using C code enhanced with ARM NEON intrinsics) to integrate the ReLU function and batch normalization as part of
the micro-kernel code.
For the ReLU function, this is straightforward as this transform is applied element-wise. For the batch
normalization, the fusion is more involved as the 2D output of the matrix multiplication needs to be
mapped into the 4D result of the convolution as part of the fused normalization.

From the implementation point of view, the specialized case of the micro-kernel with fused layers is invoked from \gemm to update matrix $C$ during  the final iteration of the loop that traverses the $k$-dimension of the problem (indexed by $p_c$); see Figure~\ref{fig:gotoblas_gemm}. For the remaining iterations of that loop, the \gemm realization invokes the regular (i.e., non-fused) micro-kernel.

The \fuse column in Table~\ref{tab:cost} shows a cost for the fused Conv2D operations of \SI{7.78}{s} only, which is even lower than that of the non-fused operations in \cache-opt. This is due to the use of the new micro-kernel, with vector intrinsics. This
reduction is even more notable if we consider that this micro-kernel not only performs the Conv2D operations but also
all the batch normalizations as well as the ReLU functions which could be fused (about half of them).
As a result, we obtain a raise in the throughput rate from \SI{12.68}{s} for \cache-opt to \SI{14.34}{s} for \fuse, which corresponds to a speedup of 1.13.
% that, for the starting \cache-opt variant, the time costs of the Conv2D
%the batch normalization
%and ReLU, 0.40 and \SI{1.04}{s} respectively, are reduced in the \fuse variant to \SI{0.47}{s} for the fused transform
%plus \SI{0.60}{s} for the ReLU transforms which cannot be fused (because they do not follow a batch normalization).
%As a result, the contribution of the batch normalization+ReLU is reduced from
%4.09\%+10.45\% = 14.54\% in \batn-opt to 4.90\%+6.30\% = 11.20\% in \fuse, and the images/s rate grows from 12.84 to 13.25 (that is, 3.1\%) for the respective variants.

\section{General Evaluation}
\label{sec:evaluation}

\subsection{Comparison with other frameworks}

The global result of the multi-step optimization process described in the previous sections shows an increase in the processing rate from the original
\SI{5.31}{images/s} with the prototype inference module derived from \pydl up to
\SI{14.34}{images/s,} which yields a global speed-up of~2.70 over the original solution.

Table~\ref{tab:inferencethroutput} compares the throughput of the optimized inference module based on \pydl with the state-of-the-art results reported in the latest release of the MLPerf benchmark (1.0)
using the Carmel processor in the NVIDIA's Jetson AGX Xavier platform.%
%The results for ArmNN (v21.02) and TFLite (v2.4.1) are extracted from the MLPerf inference webpage.
\footnote{\url{https://mlperf.org/inference-results/}}
%In these scenarios, the batch size is unknown.
%In the case of TFLite 2.4.0, the results have been obtained with the benchmark model tflite tool that is included in the TF framework.
%For comparison, on the same testbed, TensorFlow Lite (TFLite) delivers between 9 and 13 images/s,
%depending on the version (2.2.0, 2.3.0 and 2.4.0) for a batch size of 128. {\color{red} No sabemos el batch size para las versiones 2.2 y 2.3 puesto que no lo dice en el test.}
Overall, (one instance of) \pydl outperforms TFLite while being slower than the native ArmNN.
At this point though, we would like to remark that the optimization of \pydl
was achieved via high-level transformations of the original
Python code for \pydl, the development of some high-level Cython routines,
the encoding of a NEON-based micro-kernel with fused layers, and the appropriate selection of some configuration parameters.
Therefore, we believe that our techniques are quite general, yielding a moderately portable inference engine
to different platforms from NVIDIA/ARM as well as from other vendors.

% \begin{table}
% \centering
% %\begin{center}
% \resizebox{\textwidth}{!}{
% \begin{tabular}{|l||r|r|r|r|}
% \hline
% Framework            & Version       &Images/s & MLPerf &Batch size (images) \\  \hline  \hline
% TFLite               & 2.4.1(ruy)    &  13    & 1.0   & Unknown\\ \hline
% ArmNN                & 21.02(Neon)   & 18     & 1.0   & Unknown \\ \hline
% %ArmNN               & 20.08(Neon)      & 17     & 0.7   & Unknown \\ \hline
% \pydl (1 instance)   & 1.0    & 14.34  & --    & 128\\ \hline
% \pydl (4 instances)  & 1.0    & 16.40  & --    & 32 /instance\\ \hline
% %TFLite        & 2.2.0(ruy)       &  13    & 0.7   & Unknown\\ \hline
% %TFLite        & 2.3.0(ruy)       &  9     & 0.7   & Unknown\\ \hline
% \end{tabular}}
% %\end{center}
% \caption{Throughput comparison of inference frameworks using the full 8 Carmel cores of NVIDIA's Jetson AGX Xavier.}
% \label{tab:inferencethroutput}
% \end{table}

\begin{table}
 \centering
 \resizebox{\linewidth}{!}{
  \begin{tabular}{rcrcr}
   \toprule
   Framework           & Version     & Images/s        & MLPerf & Batch size (images) \\  \midrule
   TFLite              & 2.4.1 (ruy)  & 13\phantom{.00} & 1.0    & Unknown             \\
   ArmNN               & 21.02 (Neon) & 18\phantom{.00} & 1.0    & Unknown             \\
   %ArmNN               & 20.08(Neon)      & 17     & 0.7   & Unknown \\ \hline
   \pydl (1 instance)  & 1.0         & 14.34           & ---    & 128                 \\
   \pydl (4 instances) & 1.0         & 16.40           & ---    & 32/instance         \\
   \pydl (8 instances) & 1.0         & 16.56           & ---    & 16/instance         \\
   %TFLite        & 2.2.0(ruy)       &  13    & 0.7   & Unknown\\ \hline
   %TFLite        & 2.3.0(ruy)       &  9     & 0.7   & Unknown\\ \hline
   \bottomrule
  \end{tabular}
 }
 \caption{Throughput comparison of inference frameworks using the full 8 Carmel cores of NVIDIA's Jetson AGX Xavier.}
 \label{tab:inferencethroutput}
\end{table}

\subsection{Real-time inference and alternative parallelization schemes}

\begin{figure}[tbh!]
 \centering\vspace{0.1cm}
 \includegraphics[width=0.48\textwidth]{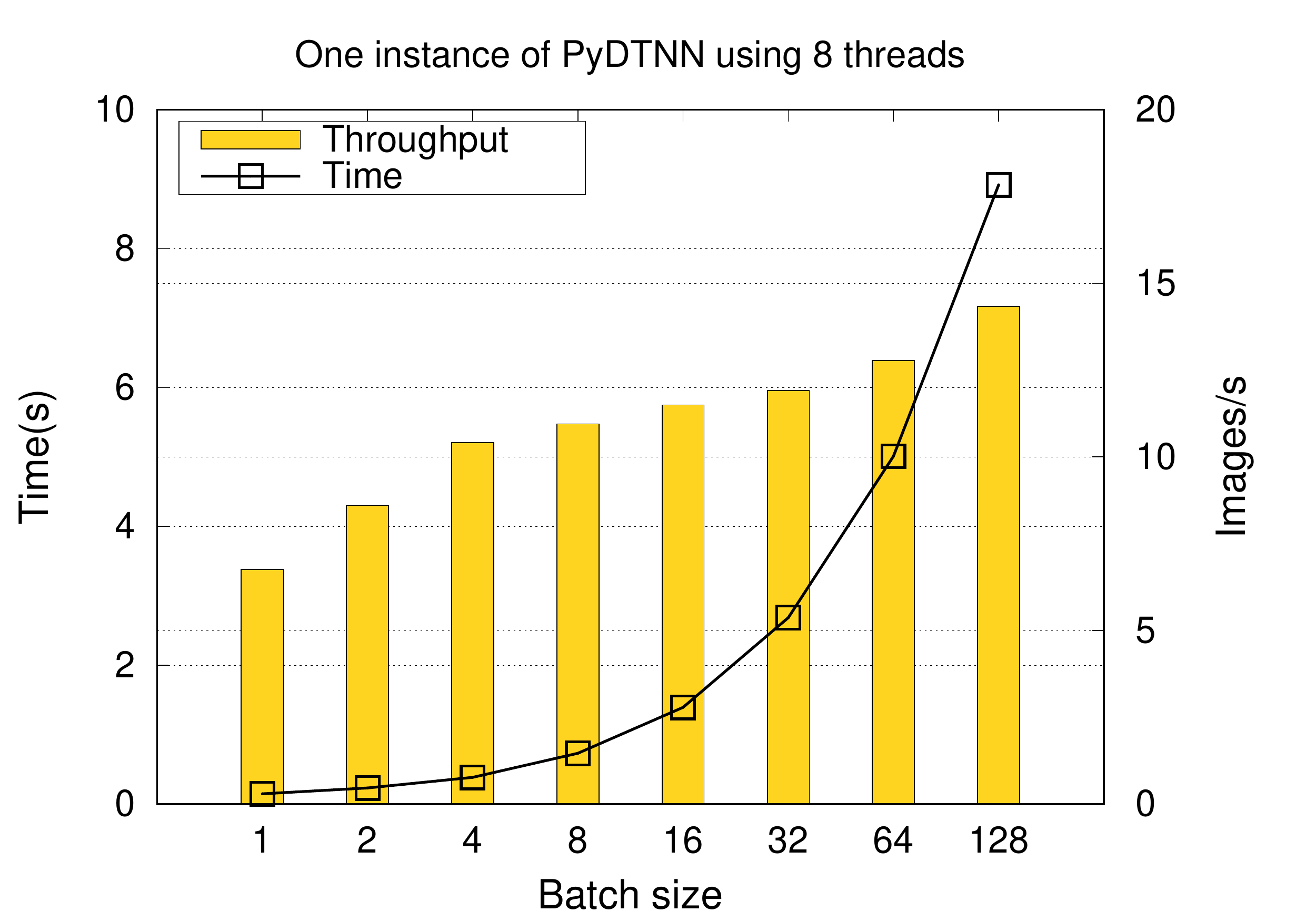}
 \includegraphics[width=0.48\textwidth]{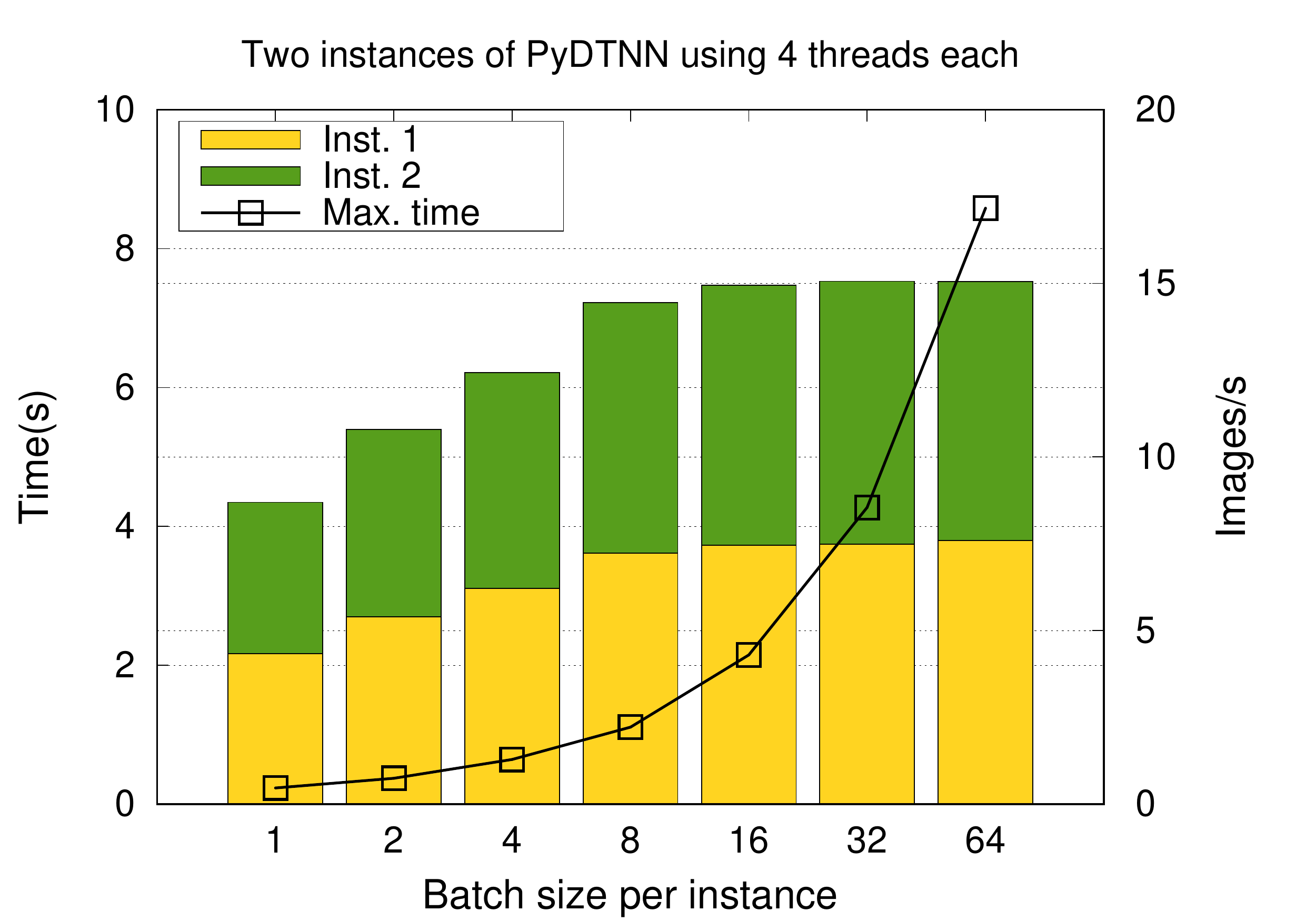}\\
 \includegraphics[width=0.48\textwidth]{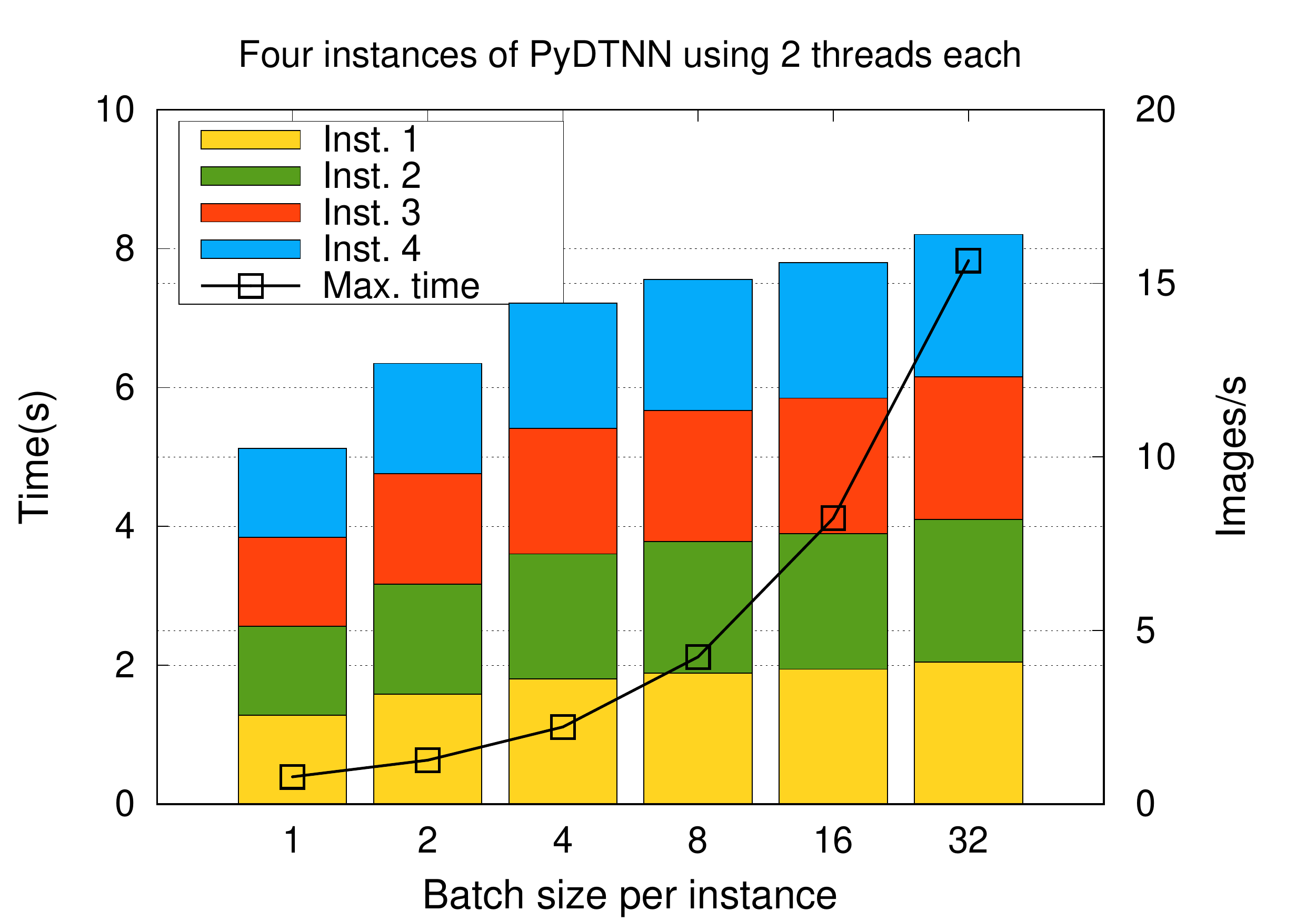}
 \includegraphics[width=0.48\textwidth]{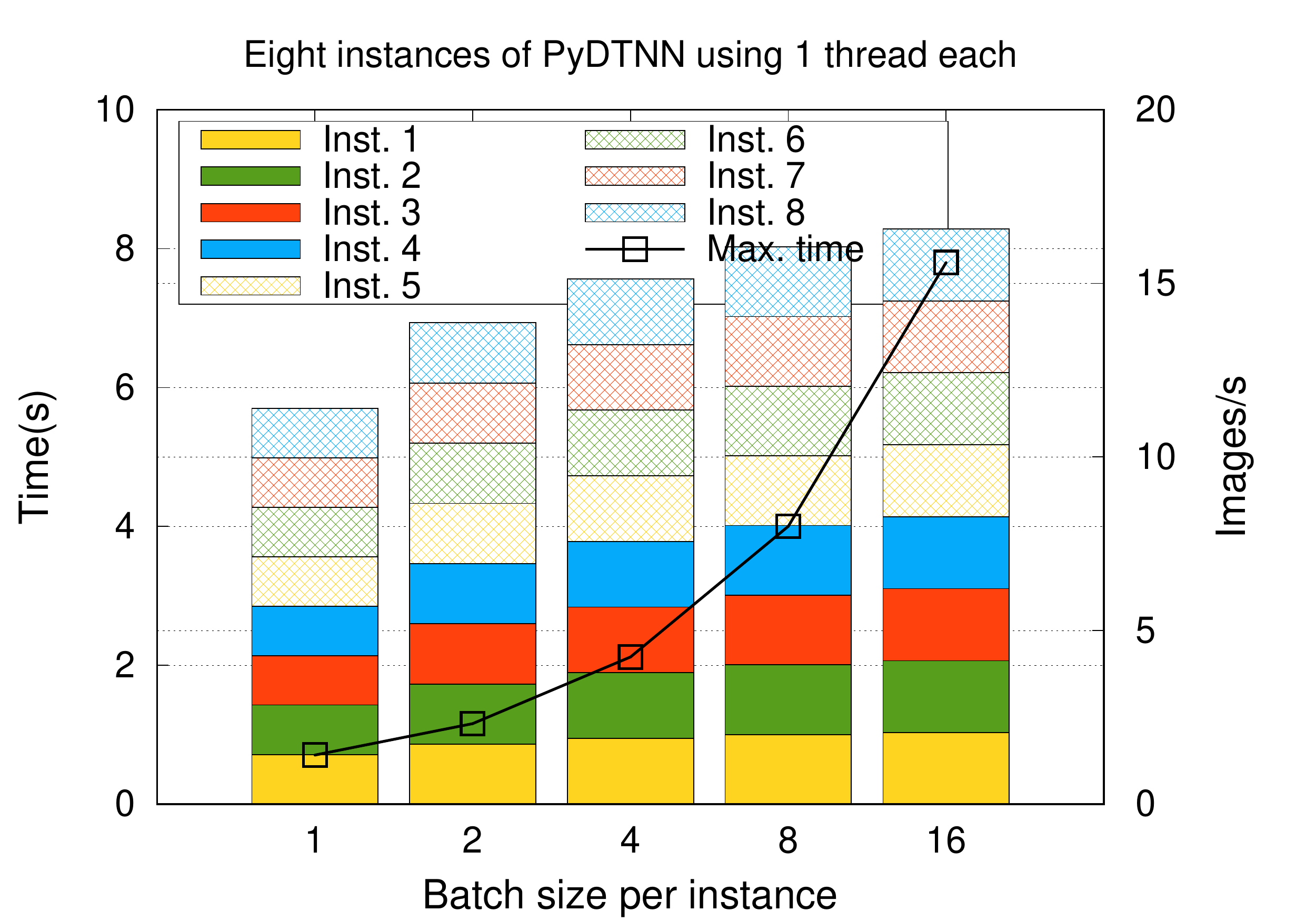}
 \caption{Maximum inference time (for all instances) and throughput (in images/s) when running 1, 2, 4, and 8 concurrent instances of
  \pydl respectively using 8, 4, 2, and 1 threads each (top-left, top-right, bottom-left, and bottom-right, respectively).}
 \label{fig:instances}
\end{figure}

In time-constrained scenarios, raw inference throughput (i.e., processed images/s) is a secondary figure-of-merit and the evaluation should be instead focused on the time-to-response.
The plot in the top-left of Figure~\ref{fig:instances} reports the time to process a batch consisting of a varying number of images, from $t=1$ to 128, using the full eight ARM
cores in the NVIDIA Jetson board. This experiment shows that increasing the batch size gradually improves the throughput
(images/s) of the inference engine, especially for small values of~$t$, but also augments the total execution time
for the batch. Thus, depending on the specific constraint determined by a given application on the time-to-response, the results identify the maximum batch size that can be used while still meeting that threshold and maximizing productivity.
For instance, an upper bound of \SI{2}{s} in the time-to-response is met for a batch with $t=16$, but not for $t=32$.

Related to the time-to-response versus throughput balance, there is an alternative parallelization scheme that
is worth investigating: concretely, instead of executing a single instance of \pydl that utilizes all the platform cores,
we can run multiple instances of the inference engine, each mapped to a distinct subset of the platform cores and working with a separate batch of images
(provided the memory capacity allows to replicate the model).
{\color{black} The rest of the plots in
Figure~\ref{fig:instances} show the results for this experiment using
2, 4, and 8 instances of \pydl, which respectively utilize 4, 2, and 1 distinct core(s) each,
thus involving the full 8 ARM cores of the platform.}
This experiment offers a few interesting insights:
\begin{enumerate}
 \item The global throughput of this alternative scheme offers higher performance than the option that runs a single instance of \pydl. Specifically, compared with the \SI{14.34}{images/s} of the conventional parallelization approach,
       running 2 concurrent instances of \pydl delivers 15 images/s, 4 concurrent instances achieves 16.4 images/s, and % and the throughput is even higher
       %when
       \textcolor{black}{scaling up to 8 instances provides \SI{16.56}{images/s,} standing rather close to 18 images/s delivered by ArmNN}. In consequence, provided a large number of images are available at an ``input inference queue,'' we may want to process them using multiple instances of the inference engine. \textcolor{black}{Also, this parallelization scheme is particularly interesting in ensemble learning,
        where the predictions from multiple DNNs trained on different initial conditions (e.g. weights initialization) are combined to reduce the variance and generalization error of the predictions. Notice that in this scenario the same batch of images is simultaneously passed for inference to the different DNNs in the ensemble.}
       % Notice that in case we have to run not a single
       % DNN model, but several distinct ones, this is a must.
 \item All instances mostly deliver the same throughput, which demonstrates that the overheads due
       to the concurrent execution of multiple concurrent instances are negligible.
       These are good news for multi-model scenarios as those arising in autonomous vehicles.
 \item For real-time scenarios, increasing the number of concurrent instances of \pydl penalizes the time-to-response
       of a single batch. For example, a batch $B_1$, {\color{black}consisting of 16 images, is processed in \SI{1.4}{s} when executed using a single instance of \pydl using the full board resources.
       In comparison, the same batch $B_1$ takes \SI{2.14}{s} to process when there are two running instances of \pydl, one working on this batch and a second one processing a different batch $B_2$. When the number of instances of
       \pydl is raised to 4, the time is almost doubled, and it takes  \SI{4.11}{s} to process $B_1$. In the latter case, when 8 instances are concurrently launched, the processing of the batch takes \SI{7.8}{s}. 
       This was about to be expected as, in the latter case,
       the amount of resources dedicated to processing that particular batch is divided by 8, 4 or 2, compared with the single-instance, 2-instance, and 4-instance execution, respectively.}
\end{enumerate}

\subsection{Energy consumption, power modes, and performance}

Energy consumption is a relevant metric for energy-constrained platforms,
such as battery-powered embedded devices.
In the following experiment, we evaluate the
energy consumption per image attained with the most significant power modes in the Jetson AGX Xavier board:
\WALL and \MAXN. In the former
case, the operating system regulates the hardware elements in the board
(in particular, the frequency of the CPU cores and the GPU)
to ensure that the system does not exceed a threshold power dissipation of \SI{30}{watts}. In practice, we observed that,
for example, this results in the processor frequency for the ARM cores being set to \SI{1.2}{GHz} when all 8 cores are utilized.
In the latter power mode, the frequency is set to \SI{2.3}{GHz} for all ARM cores.

\begin{figure}[tbh!]
 \centering\vspace{0.1cm}
 \includegraphics[width=0.482\textwidth]{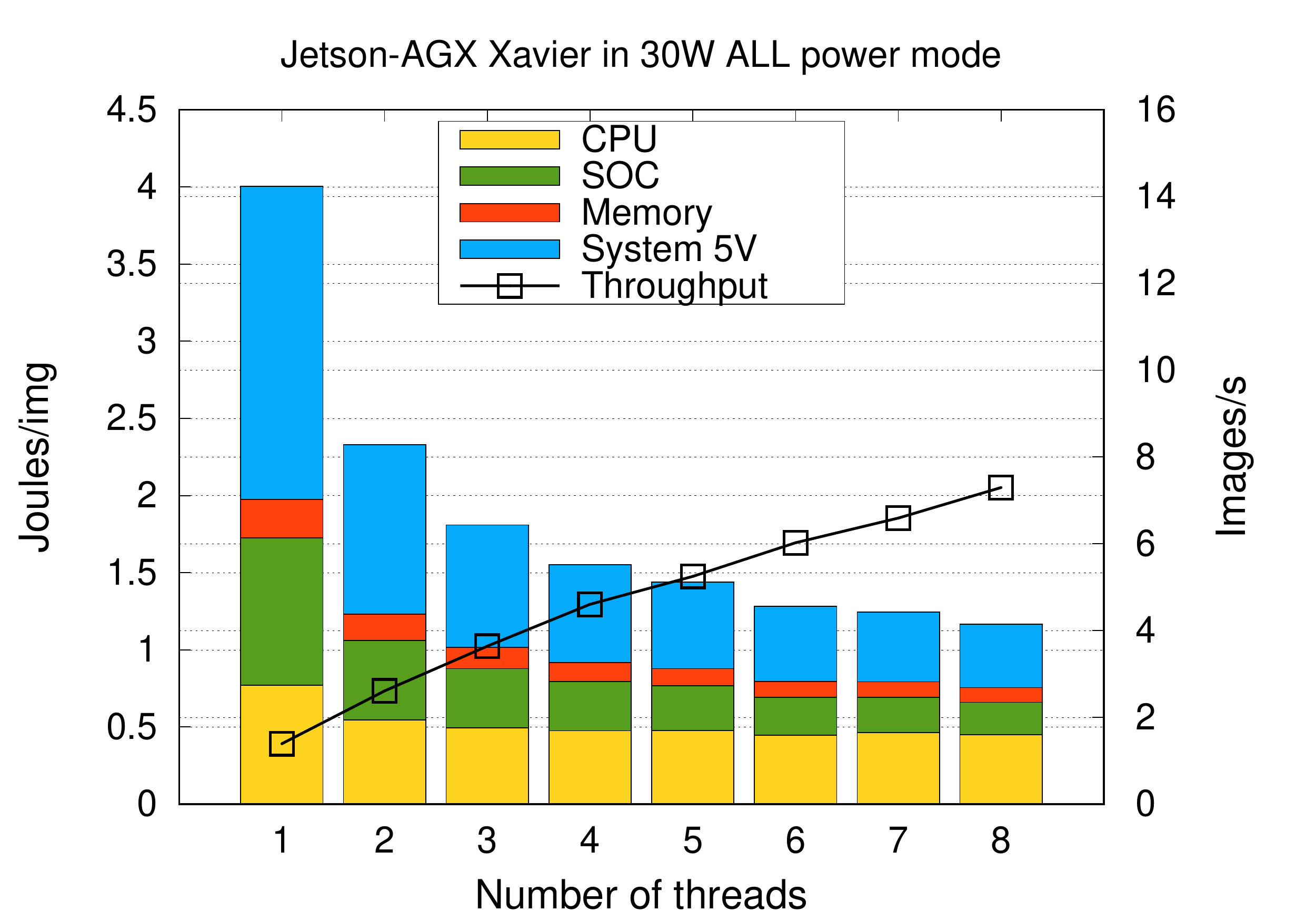}
 \includegraphics[width=0.482\textwidth]{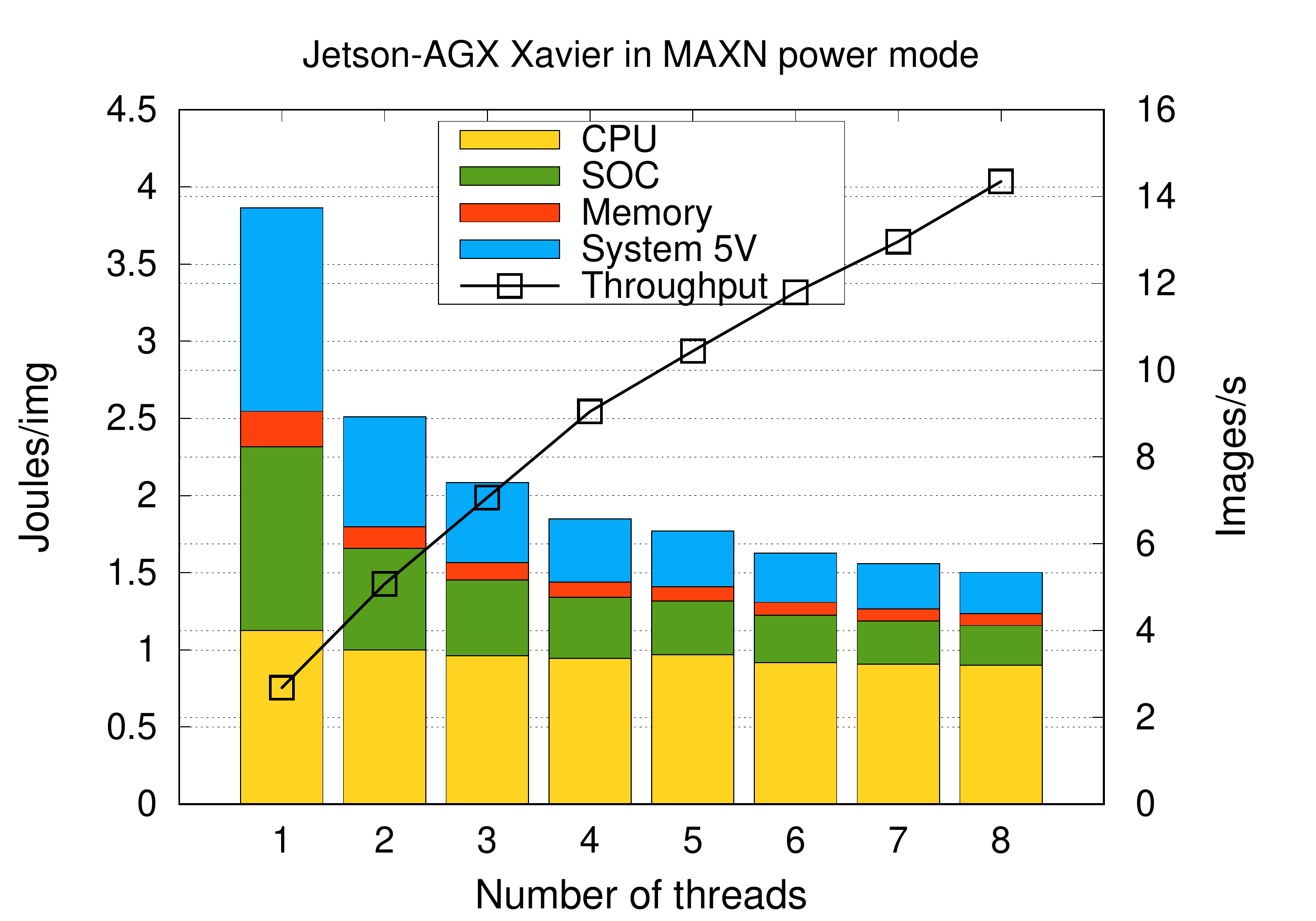}\\
 \includegraphics[width=0.482\textwidth]{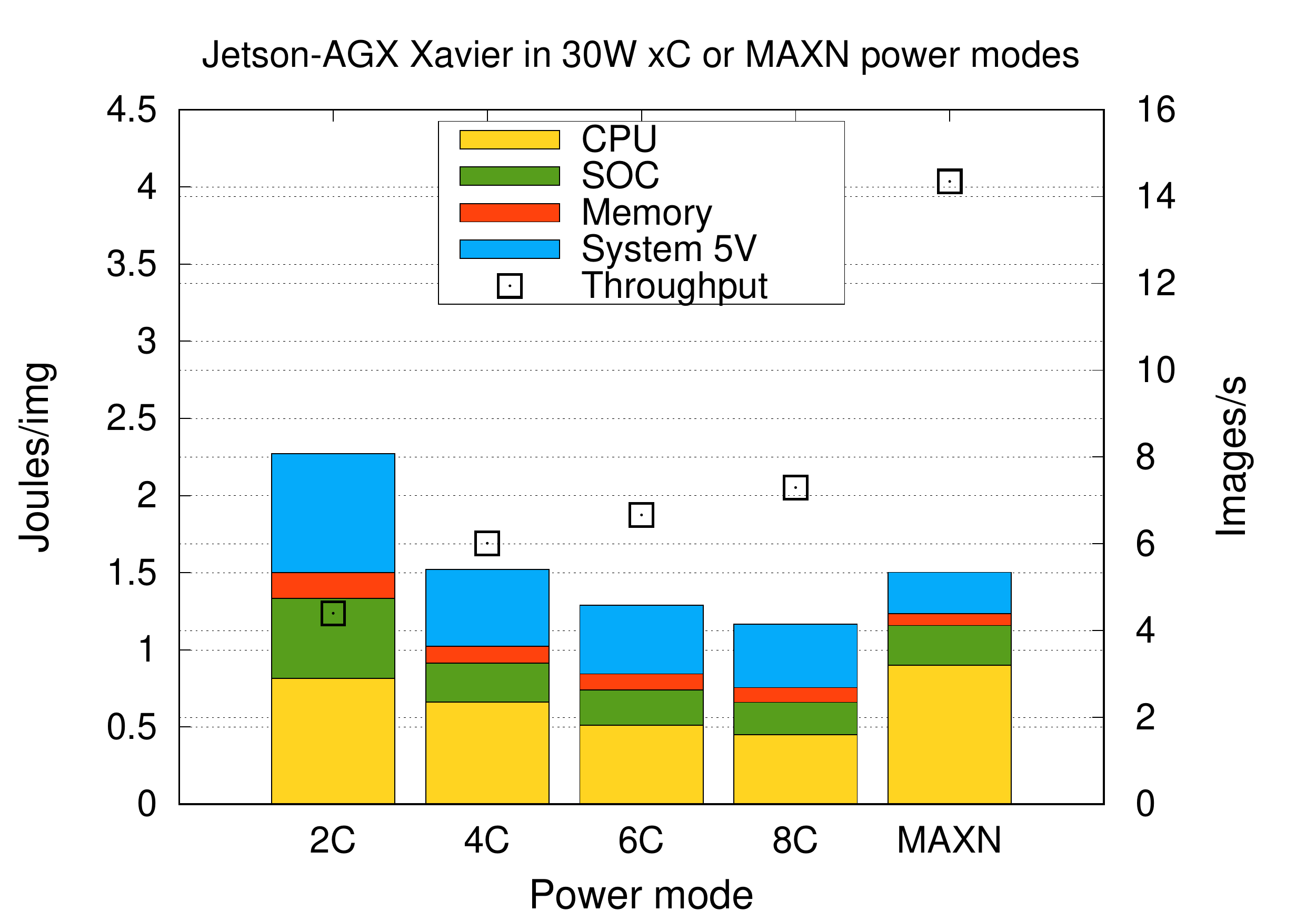}
 \caption{Inference energy consumption per image and throughput (in images/s) when running 1 instance of \pydl with the NVIDIA Jetson AGX Xavier in two power modes, \WALL and \MAXN (top-left and top-right, respectively) using 1--8 cores;
  and using the \textsf{30W xC} power modes which deactivate some of the socket cores ($8-\textsf{x}$) versus the \MAXN power mode.}
 \label{fig:energy_mxn_vs_30W}
\end{figure}

The top two plots in Figure~\ref{fig:energy_mxn_vs_30W} show the energy consumption per image
when running a single instance of \pydl that processes a batch of $t=128$ images using an increasing number of threads,
from 1 to 8, with the board operating in the
\WALL or \MAXN power modes. We can observe the significant impact of the power modes on the inference throughput, which
is roughly multiplied by two for \MAXN compared with \WALL. This is natural if we take into account the increase of
the CPU frequency which is also mostly doubled from \SI{1.2}{GHz} to \SI{2.3}{GHz}.
However, the performance boost comes at the cost of some energy efficiency loss, as the plots show that, for example,
8 threads running in \WALL consume about \SI{1.2}{joules/image} while the same cores in \MAXN
increase energy consumption to \SI{1.5}{joules/image}.

To close the analysis of energy consumption, the bottom plot in Figure~\ref{fig:energy_mxn_vs_30W} reports
the energy consumption per image and throughput when we employ the alternative
\textsf{30W xC} power mode, where \textsf{x} specifies the number of active cores while the remaining $8-\textsf{x}$ cores
are switched off. Note that this is different from the \WALL power mode as in this later configuration all
cores are turned on, though some of them may be idle (C state) because they do not intervene in the execution of the inference module. \textcolor{black}{Due to that, the CPU frequency of the active cores can be raised beyond \SI{1.2}{GHz} up to a higher clock rate, without exceeding the power budget of \SI{30}{W}.}
For instance, comparing the energy consumption (per image) of the
\textsf{30W 2C} power mode with that observed for the
execution using two cores in
\WALL power mode, we can appreciate a slightly higher consumption for the latter, which was to be expected as
idle cores still consume some energy. \textcolor{black}{Given the CPU frequency increase in \textsf{30W 2C}, the inference throughput of \SI{2.4}{images/s} using 2 cores in the \WALL mode is almost doubled to \SI{4.5}{images/s} in \textsf{30W 2C}.}

%In contrast, from the point of view of inference throughput,
%there is no difference between the two configurations.

\section{Concluding Remarks}
\label{sec:remarks}

In this paper, we have evolved \pydl, a framework for distributed parallel training of Deep Neural Networks (DNNs), to efficiently perform
%We have elaborated, described and experimentally analyzed an efficient tool for the
inference with convolutional neural networks on multicore ARM processors.
This new inference engine inherits %some of
the appealing features
of the ancestor \pydl training framework in terms of simplicity, user-friendly interface, and support for popular DNNs such
as MLPs, CNNs, ResNets, and transformers. In addition, this  inference tool
applies some general-purpose high performance techniques as well as a few architecture-specific
optimizations to deliver inference throughput that is competitive with that observed for popular frameworks, such as
TF Lite, as well as highly architecture-dependent counterparts for ARM-based processors such as ArmNN.

\textcolor{black}{The results for the multi-instance parallelization scenario demonstrate the relative importance of the batch size in the time-to-response versus raw throughput (images/s). Also, the experiments evaluating the energy consumption in the different modes reveal that a higher throughput comes at the cost of higher energy consumption. However, disabling cores helps to increase the throughput at a constant power budget.}

\section*{Acknowledgments}
This research  was partially sponsored by projects
TIN2017-82972-R of
\textit{Ministerio de Ciencia, Innovaci\'on y Universidades}
and Prometeo/2019/109 of the
\textit{Generalitat Valenciana}. Adri\'an Castell\'o was supported by the Juan de la Cierva-Formaci\'on project FJC2019-039222-I of the \emph{Ministerio de Ciencia, Innovaci\'on y Universidades}. Manuel F. Dolz was also supported by the Plan GenT project CDEIGENT/2018/014 of the \emph{Generalitat Valenciana}.

\bibliographystyle{elsarticle-num}
\bibliography{biblio,hpc,deep}

\end{document}